\documentclass[twocolumn,showpacs,preprintnumbers,amsmath,amssymb]{revtex4}

\usepackage{graphicx}% Include figure files
\usepackage{dcolumn}% Align table columns on decimal point
\usepackage{bm}% bold math

\begin{document}
\draft
\title{Bound states in the continuum in a single-level Fano-Anderson model}
\normalsize
\author{Stefano Longhi}
\address{Dipartimento di Fisica and Istituto di Fotonica e Nanotecnologie del CNR,
Politecnico di Milano, Piazza L. da Vinci 32,  I-20133 Milan,
Italy}

%\date{.}

%
\bigskip
\begin{abstract}
\noindent Bound states in the continuum (BIC) are shown to exist
in a single-level Fano-Anderson model with a colored interaction
between the discrete state and a tight-binding continuum, which
may describe mesoscopic electron or photon transport in a
semi-infinite one-dimensional lattice. The existence of BIC is
explained in the lattice realization as a boundary effect induced
by lattice truncation.
\end{abstract}


% 73.23.Ad Ballistic transport
% 03.65.Ge Solutions of wave equations: bound states
% 73.23.-b Electronic transport in mesoscopic systems

 \maketitle

\newpage

\section{Introduction.}
Since the pioneering proposal of the existence of isolated quantum
mechanical bound states embedded in the continuum, made by von
Neumann and Wigner \cite{vonNeumann29} in the study of the
one-particle Schr\"{o}dinger equation with certain spatially
oscillating attractive potentials, several theoretical and a few
experimental studies have demonstrated the existence of "bound
states in the continuum" (BIC) in a wide range of different
physical systems
\cite{Stillinger75,Stillinger77,Herrick77,Friedrich85,
Timp88,Schult90,Capasso92,Ji92,Deo94,Weber94,Kim99,Nockel99,Olendski02,Bulgakov02,Cederbaum03,
Ladron03,Sprung03,Rotter05,Ladron06,Ordonez06,Voo06}. BIC may be
found in certain atomic or molecular systems
\cite{Stillinger75,Friedrich85,Cederbaum03}, such as hydrogen atom
in a uniform magnetic field \cite{Friedrich85}, in semiconductor
superlattice structures
\cite{Stillinger77,Herrick77,Capasso92,Weber94,Sprung03}, in
mesoscopic electron transport and quantum waveguides
\cite{Timp88,Schult90,Ji92,Deo94,Kim99,Olendski02,Bulgakov02}, and
in quantum dot systems
\cite{Nockel99,Ladron03,Rotter05,Ladron06,Ordonez06}. The question
of existence of BIC has been also addressed for the famous
Fano-Anderson Hamiltonian \cite{Fano,Anderson,Mahan} (also
referred to as the Friedrichs-Lee Hamiltonian
\cite{Friedrichs,Lee,Prigogine92}), which model the process of
quantum mechanical decay of an unstable localized state coupled
with a continuum in different contexts such as atomic physics,
quantum electrodynamics, solid state and high-energy physics (see,
e.g., \cite{Mahan,Tannoudji,Knight90,Lambropoulos00,Gadzuk00}).
BIC in Fano-Anderson-like models are commonplace in case where
{\it several} (i.e. more than one) discrete states are coupled to
a common continuum; a noteworthy example of this case in
condensed-matter physics is provided, for instance, by
 quantum transport and scattering in dot molecules attached to
leads \cite{Ladron03,Rotter05,Ladron06,Ordonez06,Voo06}. In those
systems the existence of BIC is usually related to the destruction
of discrete-continuum decay channels via quantum interference
through a typical trapping mechanism. The conditions for the
existence of BIC for a general multi-level Fano-Anderson
Hamiltonian have been recently stated in Ref.\cite{Miyamoto05}; in
particular, a sufficient condition that ensures the {\it
non-existence} of BIC has been demonstrated. In case of a {\it
single} localized state embedded in and interacting with a
continuum, the existence of bound states (sometimes referred to as
"dressed bound states") has been acknowledged on many occasions
and related to threshold effects or to singularities or gaps in
the density of states of the continuum (see, e.g.,
\cite{Mahan,Lambropoulos00,Lewenstein88,Piraux90,John90,John94,Kofman94,Gaveau95}).
However, such dressed bound states  have usually an energy {\it
outside} the continuum \cite{Mahan}: Effects such as fractional
decay, population trapping and atom-photon bound states found in
several physical models describing the decay of a single discrete
level coupled to a continuum
\cite{Lambropoulos00,John94,Kofman94,Gaveau95} are in fact most of
the times related to the existence of dressed bound states with an
energy {\it outside} the continuum. Conversely, BIC have been
seldom encountered in the single-level Fano-Anderson model:
Examples of BIC involving a single localized state coupled with a
specially-structured continuum have been found in the study of
certain exactly-solvable electrodynamic models of spontaneous
emission decay with a density of modes showing a point-like gap
\cite{Kilin93,Nabiev93}, however physical realizations of such
models were not proposed \cite{Lambropoulos00}.\\
It is the aim of this work to present an exactly-solvable
Fano-Anderson Hamiltonian with a single discrete level coupled to
a continuum  by a colored interaction which supports BIC. The
proposed model is of physical relevance to certain
condensed-matter and photonic systems, since it may describe the
charge transfer dynamics of adatoms to a semi-infinite
one-dimensional lattice of quantum dots \cite{Tanaka06} or photon
tunneling dynamics in semi-infinite optical waveguide arrays or
coupled optical resonators
\cite{Stefanou98,Yariv99,Bayindir00,Fan04,Kivshar05,Longhi06}
within the tight-binding approximation. In these lattice
realizations, the tight-binding analysis allows one to simply
explain the appearance of BIC as a {\it boundary effect} due to
truncation of the lattice.\\
The paper is organized as follows. In Sec.II, the Fano-Anderson
model describing the coupling of a single localized state with a
continuum is briefly reviewed, and the conditions for the
existence of bound states either inside or outside the continuum
are presented. Section III proposes an exactly-solvable model with
colored interaction which supports BIC; exact analytical results
for the fractional decay induced by BIC are also presented.
Finally, in Sec. IV a lattice realization of the model is
presented, together with a simple physical explanation for the
existence of BIC.

\section{Bound states and decay dynamics for the single-level Fano-Anderson model}
\subsection{Basic Model}
The starting point of the analysis is provided by a standard
Fano-Anderson model describing the interaction of a discrete state
$|a\rangle$, of energy $\hbar \omega_a$, with a continuum
described by a set of continuous states $|k\rangle$ with energy $
\hbar \omega(\kappa)$ (see, for instance,
\cite{Mahan,Tannoudji,Lambropoulos00}). The Hamiltonian of the
interacting system can be written as $H=H_0+V$, where
\begin{equation}
H_0=\hbar \omega_a |a\rangle\langle a |+ \int dk \; \hbar
\omega(k) |k \rangle\langle k |
\end{equation}
is the Hamiltonian of the non-interacting discrete and continuous
states, and
\begin{equation}
V=\hbar  \int dk  \left[ v(k) |a \rangle\langle k |+v^*(k) |k
\rangle\langle a | \right]
\end{equation}
is the interaction part. Normalization has been assumed such that
$\langle a | a \rangle=1$, $\langle a | k \rangle=0$ and $\langle
k' | k \rangle= \delta(k-k')$. If we expand the wave function
$|\psi\rangle$ of the system  as $|\psi\rangle=c_a(t)|a\rangle+
\int dk c(k,t)|k \rangle$, the expansion coefficients $c_a(t)$ and
$c(k,t)$ satisfy the coupled-mode equations
\begin{eqnarray}
i \dot{c}_a(t) & = & \omega_a c_a+ \int dk v(k) c(k,t) \\
i \dot{c}(k,t) & = & \omega(k) c(k,t)+ v^*(k) c_a(t),
\end{eqnarray}
where the dot indicates the derivative with respect to time.
Typically, we assume that the frequency $\omega (k)$ of continuous
states spans a finite interval (a band) $\omega_1 < \omega <
\omega_2$ for the allowed values of the continuous variable $k$,
and that the frequency $\omega_a$ of the discrete level is
embedded in the continuum, i.e. that $\omega_1 < \omega_a <
\omega_2$.

\subsection{Bound states}
The eigenstates $| \psi_E\rangle$ of $H$ corresponding to the
eigenvalue $E= \hbar \Omega$ are obtained from the eigenvalue
equation $H | \psi_E\rangle = \hbar \Omega | \psi_E\rangle$. After
introduction of the density of states $\rho(\omega)=
\partial k / \partial \omega$, from Eqs.(1) and (2) it follows
that the eigenfrequencies $\Omega$ are found as the eigenvalues of
the system
\begin{eqnarray}
\Omega c_a & = & \omega_a c_a+ \int_{\omega_1}^{\omega_2} d \omega \sqrt{\rho(\omega)}v(\omega) \tilde{c}(\omega) \\
\Omega \tilde{c}(\omega) & = & \omega \tilde{c}(\omega)+
\sqrt{\rho(\omega)} v^*(\omega)c_a,
\end{eqnarray}
where we have set $\tilde{c}(\omega)=\sqrt{\rho(\omega)}
c(\omega)$. As the continuous spectrum of $H$ is the same as that
of $H_0$, isolated eigenvalues corresponding to bound states may
or may not occur for $H$. The energy $E$ corresponds to a bound
state of $H$ provided that $| \psi_E\rangle$ is square integrable.
This implies
\begin{equation}
\| \psi_E \|^2=|c_a|^2+ \int_{\omega_1}^{\omega_2} d \omega
|\tilde{c}(\omega)|^2 < \infty.
\end{equation}
We have to distinguish two cases.\\

(i) {\it Bound states outside the continuum}. This is the most
common case, which has been studied on several occasions. A bound
state with energy $ \hbar \Omega$ outside the continuum exists
provided that a root of the equation
\begin{equation}
\Omega-\omega_a= \Delta (\Omega)
\end{equation}
can be found outside the band $(\omega_1, \omega_2)$, where we
have set
\begin{equation}
\Delta (\Omega)=\int_{\omega_1}^{\omega_2}d \omega
\frac{\rho(\omega) |v(\omega)|^2}{\Omega-\omega}.
\end{equation}
The conditions for the existence of bound states of this kind have
been extensively investigated in the literature (see, e.g.,
\cite{Lambropoulos00,Kofman94,Gaveau95}). For instance, Eq.(8)
admits always two solutions outside the interval $(\omega_1,
\omega_2)$ whenever $\rho(\omega)|v(\omega)|^2$ does not vanish at
the edge of the band, since in this case $\Delta(\Omega)$ diverges
to $\mp \infty$ as $
\Omega \rightarrow \omega_{1,2}^{\mp}$.\\
\\
(ii) {\it Bound states inside the continuum.} As shown in
Ref.\cite{Miyamoto05}, a bound state at frequency $\Omega$ inside
the continuum may exist provided that the following two conditions
are simultaneously satisfied
\begin{equation}
|v(\Omega)|^2 \rho(\Omega)=0 \; \; , \; \;
\Omega-\omega_a=\Delta(\Omega)
\end{equation}
Additionally, $v(\omega) \sqrt{\rho(\omega)}$ should vanish as
$\omega \rightarrow \Omega$ at least as $\sim (\omega-\Omega)$ in
order to ensure a finite norm [Eq.(7)]. The first equation in (10)
can be satisfied for either $\rho(\Omega)=0$ or $v(\Omega)=0$. The
former case, which corresponds to a point-like gap in the density
of states inside the band, has been previously considered for some
special density of state profiles \cite{Kilin93,Nabiev93}, which
however do not seem to have simple physical realizations
\cite{Lambropoulos00}. The latter case, $v(\Omega)=0$, implies
that the discrete state $|a \rangle$ does not interact with the
continuous state of frequency $\Omega$, and thus implies a
"colored" interaction profile $v(\omega)$ with one zero at
$\omega=\Omega$. However, at such a frequency the additional
condition $\omega_a=\Omega-\Delta(\Omega)$ must be simultaneously
satisfied, which means that BIC may exist solely at a prescribed
energy $\hbar \omega_a$ of the level $|a\rangle$.

\subsection{Decay dynamics}
Consider now the decay dynamics of the unstable state $|a\rangle$
embedded in the continuum. This corresponds to solve Eqs.(3) and
(4) with the initial conditions $c_a(0)=1$ and $c(k,0)=0$, which
can be done by e.g. a Laplace transformation or a Green's function
analysis (see, for instance,
\cite{Mahan,Tannoudji,Lambropoulos00}). Indicating by $
\hat{c}_a(s)=\int_{0}^{\infty} dt c_a(t) \exp(-st)$ [${\rm
Re}(s)>0$] the Laplace transform of $c_a(t)$, from Eqs.(3) and (4)
one obtains
\begin{equation}
\hat{c}_a(s)= \frac{i}{is-\omega_a-\Sigma(s)} ,
\end{equation}
and then, after inversion
\begin{equation}
c_a(t)=\frac{1}{2 \pi } \int_{0^+ -i \infty}^{0^+ +i \infty} ds
\frac{\exp(st)}{is-\omega_a-\Sigma(s)} ,
\end{equation}
where $\Sigma(s)$ is the self-energy, given by
\begin{equation}
\Sigma(s)=\int dk
\frac{|v(k)|^2}{is-\omega(k)}=\int_{\omega_1}^{\omega_2} d \omega
\frac{ \rho( \omega) |v(\omega)|^2}{is-\omega}.
\end{equation}
Possible poles on the imaginary axis of $ \hat{c}_a(s)$ correspond
to bound states of $H$ and are responsible for fractional decay of
the amplitude $c_a(t)$. In fact, using the property
\begin{equation}
\Sigma(s=-i \omega \pm 0^+)=\Delta (\omega) \mp i \pi \rho(\omega)
|v(\omega)|^2
\end{equation}
with
\begin{equation}
\Delta(\omega)=\mathcal{P} \int_{\omega_1}^{\omega_2} d \omega '
\frac{\rho(\omega ')|v(\omega ')|^2}{\omega-\omega '}
\end{equation}
 the poles $s_p=-i\Omega$ of $\hat{c}_a(s)$ satisfy the
conditions $\Omega-\omega_a=\Delta(\Omega)$ and $\rho(\Omega)
|v(\Omega)|^2=0$, i.e. they are located in correspondence of the
bound states (either outside or inside the continuum) of $H$. In
absence of poles (i.e. of bound states of $H$), $c_a(t)$ decays to
zero, whereas in presence of poles Eq.(12) can be written as the
sum of a contour (decaying) integral plus the (non-decaying) pole
contributions.

\section{BIC in a Fano-Anderson model with a colored interaction: an exactly-solvable model}
In this section we present an exactly-solvable Fano-Anderson
model, describing the interaction of a single discrete state with
a continuum, which admits of BIC. Precisely, we assume the
following colored interaction function
\begin{equation}
v(k)=\sqrt{\frac{2}{\pi}} \kappa_a \sin(n_0 k)
\end{equation}
and the following tight-binding dispersion curve for the band of
continuous states
\begin{equation}
\omega(k)=-2 \kappa_0 \cos k
\end{equation}
where $0 \leq k \leq \pi$, $\kappa_0$ and $\kappa_a$ are positive
real parameters, and $n_0$ is a positive and nonvanishing integer
number, i.e. $n_0=1,2,3,...$. A physical realization of this model
will be described in the next section. We note that this model is
a generalization of the well-known tight-binding Fano-Anderson
model with a constant interaction coupling $v(k)= {\rm const}$,
which is known to show bound states {\it outside} the continuum
\cite{Mahan}. Note also that for this band model the density of
states, given by
\begin{equation}
\rho(\omega)=\frac{\partial k}{\partial \omega}=\frac{1}{\sqrt{4
\kappa_{0}^2-\omega^2}},
\end{equation}
shows van-Hove singularities at the band edges $\omega=\pm 2
\kappa_0$. For a constant coupling \cite{Mahan}, these
singularities are responsible for the existence of two bound
states outside the band from either sides. However, for the
colored coupling considered in our case [Eq.(16)] one has
\begin{equation}
G(\omega) \equiv \rho(\omega)|v(\omega)|^2=\frac{2
\kappa_{a}^2}{\pi \sqrt{4 \kappa_{0}^2-\omega^2}} \sin^2 \left[
n_0 {\rm acos} \left( \frac{\omega}{2 \kappa_0} \right)  \right]
\end{equation}
which vanishes at the band edge. From Eqs.(13), (16) and (17), the
self-energy $\Sigma(s)$ can be calculated in an exact form and
reads
\begin{widetext}
\begin{equation}
\Sigma(s)=\frac{2 \kappa_{a}^2}{\pi} \int_{0}^{\pi} dk
\frac{\sin^2 (n_0 k)}{is+2 \kappa_0 \cos k}=-\frac{i
\kappa_{a}^2}{\sqrt{s^2+4 \kappa_{0}^2}} \left[1-\left( \frac{i
\sqrt{s^2+4 \kappa_{0}^2}-is}{2 \kappa_0} \right)^{2n_0} \right].
\end{equation}
\end{widetext}
Using Eq.(14), the following expression for $\Delta(\omega)={\rm
Re}[\Sigma(s=-i \omega \pm 0^+)]$ can be then derived
\begin{widetext}
\begin{eqnarray}
\Delta(\omega)= \left\{
\begin{array}{ll}
-\frac{\kappa_{a}^2}{\sqrt{\omega^2-4 \kappa_{0}^2}} \left[
1-\left( \frac{\sqrt{\omega^2-4 \kappa_{0}^2}+\omega}{2 \kappa_0}
\right)^{2 n_0} \right] & , \;  \omega < - 2 \kappa_0 \\
\frac{\kappa_{a}^2}{\sqrt{4 \kappa_{0}^2-\omega^2}} \sin \left[ 2
n_0 {\rm acos} \left( \frac{\omega}{2 \kappa_0} \right) \right] &
, \;
-2 \kappa_0 < \omega <2 \kappa_0 \\
\frac{\kappa_{a}^2}{\sqrt{\omega^2-4 \kappa_{0}^2}} \left[
1-\left( \frac{\sqrt{\omega^2-4 \kappa_{0}^2}-\omega}{2 \kappa_0}
\right)^{2 n_0} \right] & , \; \omega > 2 \kappa_0
\end{array}
\right.
\end{eqnarray}
\end{widetext}
 The behavior of $G(\omega)  \equiv
\rho(\omega)|v(\omega)|^2$ [Eq.(19)] and $\Delta(\omega)$
[Eq.(21)] for increasing values of $n_0$ is shown in Fig.1.
\begin{figure}
\includegraphics[scale=0.4]{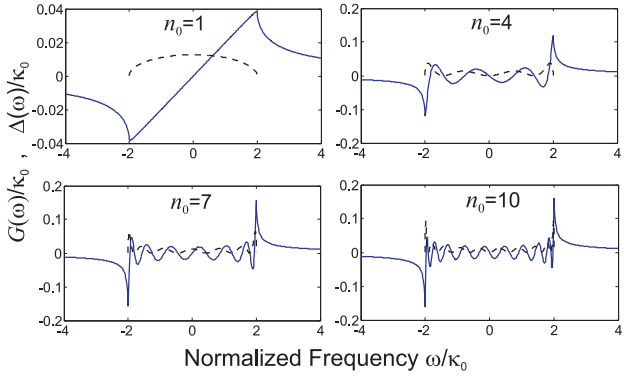} \caption{
(color online) Behavior of $G(\omega)=\rho(\omega) |v(\omega)|^2$
(dotted curves) and $\Delta(\omega)$ (solid curves), normalized to
$\kappa_0$, versus normalized frequency $\omega / \kappa_0$ for
$(\kappa_a/ \kappa_0)=0.2$ and for increasing values of integer
$n_0$.}
\end{figure}
Note the oscillatory behavior of both $G(\omega)$ and
$\Delta(\omega)$, with the existence of $(2n_0-1)$ zeros of
$\Delta(\omega)$ at $\omega_l=-2 \kappa_0 \cos [ l \pi /(2n_0)] $
($l=1,2,...,2n_0-1$) and of $(n_0+1)$ zeros of $G(\omega)$ at
$\omega_m=-2 \kappa_0 \cos (m \pi /n_0) $ ($m=0,1,2,...,n_0$). We
can then specialize the general
results of Sec.II to the present model.\\
\\
(i) {\it Bound states outside the continuum.} At most two bound
states at frequency $\Omega$ outside the band $(-2 \kappa_0, 2
\kappa_0)$ from either sides may exist. Precisely, a bound state
at frequency $\Omega> 2 \kappa_0$ exists provided that $2
\kappa_0-\omega_a< \Delta(2 \kappa_0)$, i.e. $\omega_a>2
\kappa_0-\kappa_{a}^2 n_0 / \kappa_0$, whereas a bound state at
frequency $\Omega< -2 \kappa$ does exist for $-2
\kappa_0-\omega_a> \Delta(-2 \kappa_0)$, i.e. for $\omega_a<-2
\kappa_0+\kappa_{a}^2 n_0 / \kappa_0$. Therefore, if the frequency
$\omega_a$ of the discrete level lies inside the interval
\begin{equation}
-1+\frac{\kappa_{a}^2 n_0}{2 \kappa_{0}^2}< \frac{\omega_a}{2
\kappa_0}<1-\frac{\kappa_{a}^2 n_0}{2 \kappa_{0}^2}
\end{equation}
no bound states outside the continuum do exist. For a given value
of $n_0$, Eq.(22) is satisfied for a sufficiently small value of
the coupling $\kappa_{a} / \kappa_0$ and provided that the
frequency $\omega_a$ is not too close to the band edges. In the
following, it will be assumed that no bound states exist outside
the continuum.\\
\\
(ii) {\it Bound states inside the continuum.} Following the
results of Sec.II.B, one BIC for the model expressed by Eqs.(16)
and (17) does exist at $\Omega=\omega_a$ for any $n_0 \geq 2$,
provided that the frequency $\omega_a$ of the discrete level
assumes one of the following $(n_0-1)$ allowed values:
 \begin{equation}
 \omega_a=-2 \kappa_0 \cos (m \pi
/n_0) \; \; (m=1,2,...,n_0-1).
 \end{equation}
\\
(iii) {\it Decay dynamics and fractional decay due to BIC.}
Suppose that $H$ admits of one BIC but no bound states outside the
continuum, i.e. that Eqs.(22) and (23) are simultaneously
satisfied. The decay law for $c_a(t)$ is given by the inverse
Laplace transform Eq.(12). The Bromwich integration path in
Eq.(12) can be deformed into the contour $\sigma$ shown in Fig.2,
where $\hat{c}_a(s)$ is always calculated on the first Riemannian
sheet. The integral then comprises the pole contribution at
$s_p=-i\omega_a$, which arises from the semi-circles surrounding
the pole, and the principal-value integral of $[1/(2 \pi i)]
\hat{c}_a(s=-i\omega \pm 0^+ ) \exp(-i \omega t)$ along the
interval $-2 \kappa_0<\omega<2 \kappa_0$ of the imaginary axis
from the two sides ${\rm Re}(s)=\pm 0^+$ of the branch cut, i.e.
\begin{equation}
c_a(t)=c_{pole}(t)+c_{decay}(t).
\end{equation}
Using Eqs.(12), (14) (19) and (21), after some straightforward
calculations one then obtains:
\begin{equation}
c_{pole}(t)=\frac{\exp(-i \omega_a t)}{1+\frac{n_0}{2} \left(
\frac {\kappa_a}{\kappa_0} \right)^2 \left[ 1-\left(
\frac{\omega_a}{2 \kappa_0} \right)^2 \right]^{-1/2}}
\end{equation}
for the pole contribution (non-decaying term), and
\begin{widetext}
\begin{eqnarray}
c_{decay}(t) & = & \frac{1}{2 \pi} \left(
\frac{\kappa_a}{\kappa_0} \right)^2 \mathcal{P} \int_{0}^{\pi} dk
\times \\
& \times &  \; \frac{ \sin^2 (n_0 k) \exp(2i \kappa_0 t \cos k) }{
\left[\omega_a /(2 \kappa_0)+\cos k-(\kappa_a / 2 \kappa_0)^2
\sin(2 n_0 k)/ \sin k \right]^2+ [\kappa_{a}^2/(2 \kappa_{0}^2)]^2
\sin^4 (n_0 k)/ \sin^2 k} \nonumber
\end{eqnarray}
\end{widetext}
for the decay term. The existence of a BIC is thus responsible for
a fractional decay of the amplitude $c_a(t)$. Such a fractional
decay is different from the most common one encountered in other
single-level Fano-Anderson models (see, e.g., \cite{Gaveau95})
since in our model the fractional decay is due to the existence of
a BIC. Note also that, if no bound states exist, i.e. if Eq.(22)
is satisfied but one of the resonance conditions (23) is not
satisfied, the amplitude $c_a(t)$ fully decays toward zero. In
this case, one simply has $c_a(t)=c_{decay}(t)$, where
$c_{decay}(t)$ is given again by Eq.(26) in which the principal
value of the integral may be omitted.
\begin{figure}
\includegraphics[scale=0.45]{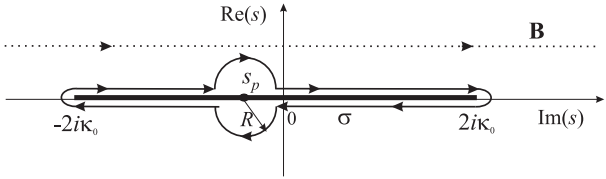} \caption{
Deformation of the Bromwich path B (dotted line) for inverse
Laplace transformation. The bold solid curve is the branch cut,
whereas the deformed path is represented by the solid closed
contour $\sigma$ surrounding the branch cut. The BIC corresponds
to the pole $s_p=-i \omega_a$ of $\hat{c}_a(s)$ on the imaginary
axis internal to the branch cut, which is surrounded by two
semi-circles whose radius $R$ tends to zero.}
\end{figure}
It is worth considering the limit $n_0 \rightarrow \infty$ for the
model expressed by Eqs.(16) and (17). In this case, in the
integral on the second right term of Eq.(20) the
rapidly-oscillating function $\sin^2 (n_0 k)$ can be replaced by
its cycle-averaged value $1/2$, i.e. one can approximately write
\begin{equation}
\Sigma(s)   \simeq  \frac{\kappa_{a}^2}{\pi} \int_{0}^{\pi} dk
\frac{1}{is+2 \kappa_0 \cos k}=\frac{-i \kappa_{a}^2}{\sqrt{s^2+4
\kappa_{0}^2}}
\end{equation}
which from Eq.(14) yields
\begin{equation}
G(\omega) \simeq \frac{\kappa_{a}^2}{\pi \sqrt{4
\kappa_{0}^2-\omega^2}}
\end{equation}
and
\begin{eqnarray}
\Delta(\omega)= \left\{
\begin{array}{ll}
-\frac{\kappa_{a}^2}{\sqrt{\omega^2-4 \kappa_{0}^2}}  & \omega < - 2 \kappa_0 \\
0 &
-2 \kappa_0 < \omega <2 \kappa_0 \\
\frac{\kappa_{a}^2}{\sqrt{\omega^2-4 \kappa_{0}^2}} & \omega > 2
\kappa_0
\end{array}
\right.
\end{eqnarray}
Note that the above expressions for $G(\omega)$ and
$\Delta(\omega)$ correspond to the limit of a tight-binding
Fano-Anderson model with an uncolored interaction (see, e.g.,
Ref.\cite{Mahan}, pp. 283-285), i.e. to a flat interaction
function $v(k) \simeq \kappa_a / \sqrt \pi$. In this case it is
knwon \cite{Mahan} that BIC do not exist at any value of
$\omega_a$, whereas two bound states outside the continuum are
always found. The physical explanation of the disappearance of BIC
in the $n_0 \rightarrow \infty$ limit will be discussed in the
next section.

\section{A tight-binding lattice realization of the colored Fano-Anderson model}
In this section we propose a simple and noteworthy physical
realization of the Fano-Anderson model with colored interaction
discussed in the previous section, which may describe either
electron or photon transport phenomena in condensed-matter or
photonic tight-binding lattices. Since $k$ varies in the range
$0<k< \pi$, we can expand $c(k,t)$ as a Fourier series of sine
terms solely according to
\begin{equation}
c(k,t)=-\sqrt{\frac{2}{\pi}} \sum_{n=1}^{\infty} c_n(t) \sin(nk)
\end{equation}
where the time-dependent coefficients $c_n$ are given by
\begin{equation}
c_n(t)=-\sqrt{\frac{2}{\pi}} \int_{0}^{\pi} dk \; c(k,t) \sin(nk).
\end{equation}
Taking into account that
\begin{equation}
\int_{0}^{\pi} dk \; \sin(nk) \sin(m k)= \frac{\pi}{2}
\delta_{n,m}
\end{equation}
($n,m \geq 1$), from Eqs.(4), (16), (17) (30) and (31) the
equations of motion for the coefficients $c_n$ can be readily
derived and read
\begin{eqnarray}
i \dot c_n & = & -\kappa_0(c_{n+1}+c_{n-1})-\kappa_a c_a
\delta_{n,n_0} \; \; (n \geq 2) \\
i \dot c_1 & = & -\kappa_0 c_{2}-\kappa_a c_a \delta_{n_0,1}.
\end{eqnarray}
The equation for $c_a$ [Eq.(3)] then reads
\begin{equation}
i \dot c_a  = \omega_a c_a-\kappa_a c_{n_0}.
\end{equation}
In their present form, Eqs.(33-35) can be derived from the
tight-binding Hamiltonian
\begin{eqnarray}
H_{TB} & = & - \hbar \sum_{n=1}^{\infty} \kappa_0 (|n \rangle
\langle n+1|+|n+1 \rangle  \langle n |) +\hbar \omega_a |a \rangle
\langle a|+ \nonumber \\
& & -\hbar \kappa_a (|a \rangle  \langle  n_0 |+|n_0 \rangle
\langle a |)
\end{eqnarray}
which describes the interaction of the localized state $|a\rangle$
with the $n_0$-th site of a semi-infinite one-dimensional
tight-binding lattice in the nearest-neighbor approximation (see
Fig.3). The tight-binding model expressed by Eq.(36) has been used
to study transport phenomena in different physical systems,
including photon tunneling dynamics in coupled optical waveguides
\cite{Longhi06} or in coupled photonic cavities
\cite{Stefanou98,Yariv99,Fan04,Kivshar05}, charge transfer of
adatoms to a one-dimensional lattice of quantum dots
\cite{Tanaka06}, or decay of the polarization in spin chains
\cite{Fiori06}. For instance, Eq.(36) may describe charge transfer
between an adatom localized state and a one-dimensional miniband
associated with a quantum dot array, the adatom being attached to
the semiconductor quantum-dot array surface \cite{Tanaka06}.
\begin{figure}
\includegraphics[scale=0.6]{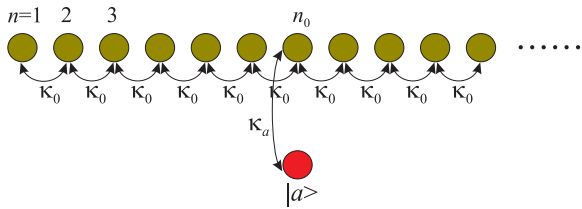} \caption{
(color online) Schematic of a localized state $|a\rangle$ coupled
to a tight-binding semi-infinite lattice which realizes the
Hamiltonian model (36). $\kappa_0$ is the hopping amplitude
between adjacent sites in the lattice, whereas $\kappa_a$ is the
hopping amplitude between the localized state $|a\rangle$ and the
site $|n_0\rangle$ of the lattice.}
\end{figure}
It should be noted that these previous models considered either a
semi-infinite tight-binding chain with a boundary defect
\cite{Longhi06,Fiori06} corresponding to the special case $n_0=1$,
where however no BIC exist, or to an infinite array
\cite{Mahan,Tanaka06}, i.e. to $n_0 \rightarrow \infty$
corresponding to a non-colored interaction, where again no BIC
exist. However, as shown in the previous section, for a finite
value of $n_0$ larger than one, i.e. by considering a
semi-infinite lattice in which a defect state interacts with a
lattice site $|n_0\rangle$ near (but not at) the boundary, BIC at
certain frequencies $\omega_a$ do exist according to Eq.(23). In
the tight-binding representation (36) the existence of BIC has a
simple physical explanation which is related to a {\it boundary
effect} of the semi-infinite lattice: BIC correspond to localized
states in the lattice with $c_n=0$ for $n \geq n_0$. In fact, let
us look for a solution to Eqs.(33-35) of the form $c_n=\bar{c}_n
\exp(-i \Omega t)$, $c_a=\bar{c}_a \exp(-i \Omega t)$, with $
\bar{c}_n=0$ for $n \geq n_0$. From Eq.(35) and Eq.(33) with
$n=n_0$ it then follows that $\Omega=\omega_a$ and
$\bar{c}_a=-(\kappa_0 / \kappa_a) \bar{c}_{n_0-1}$, whereas after
setting $\mathbf{c} \equiv
(\bar{c}_1,\bar{c}_2,....,\bar{c}_{n_0-1})^T$ from Eq.(34) and
from Eq.(33) with $n \leq n_{0}-1$ one obtains that $\mathbf{c}$
and $\Omega$ are the eigenvectors and corresponding eigenvalues of
the $(n_0-1) \times (n_0-1)$ matrix
\begin{equation}
\mathcal{M}= \left(
\begin{array}{cccccccc}
0 & -\kappa_0 & 0 & 0 & ... & 0 & 0 & 0 \\
-\kappa_0 & 0 & -\kappa_0 & 0 & ... & 0 & 0 & 0 \\
0 & -\kappa_0 & 0 & -\kappa_0 & ... & 0 & 0 & 0 \\
.. & .. & .. & .. & .. & .. & .. & .. \\
0 & 0 & 0 & 0 & ... & -\kappa_0 & 0 & -\kappa_0 \\
0 & 0 & 0 & 0 & ... & 0 & -\kappa_0 & 0 \\
\end{array}
\right),
\end{equation}
\begin{figure}
\includegraphics[scale=0.5]{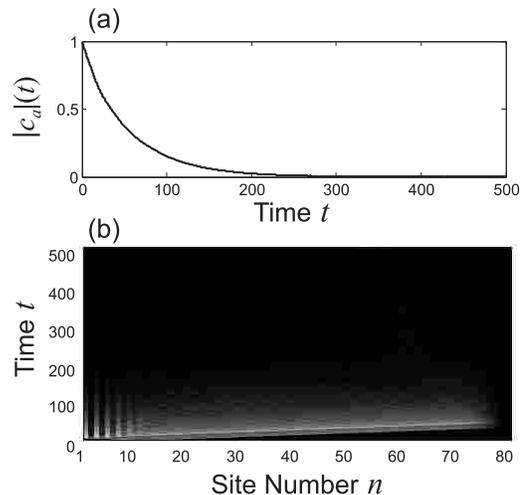} \caption{
(a) Decay dynamics of the amplitude $|c_a(t)|$ as obtained by
numerical analysis of Eqs.(33-35) for parameter values
$\kappa_0=1$, $\kappa_a=0.2$, $n_0=12$, and for $\omega_a=0.15$,
corresponding to the absence of BIC. In (b) the temporal evolution
of the amplitudes $|c_n(t)|$ at the lattice sites is also shown on
a grey-scale plot.}
\end{figure}
i.e. $\mathcal{M} \mathbf{c}=\omega_a \mathbf{c}$. Diagonalization
of the matrix $\mathcal{M}$ yields for the eigenvalues the
following expression
\begin{equation}
\Omega_m=-2 \kappa_0 \cos(m \pi / n_0) \; \; (m=1,2,...,n_0-1)
\end{equation}
with corresponding eigenvectors $\bar{c}_{n}^{(m)}=\sin ( m \pi n
/ n_0)$. Note that the values of $\Omega_m$ given by Eq.(38) are
precisely the resonance frequencies for the existence of BIC found
in the previous section [see Eq.(23)].
\begin{figure}
\includegraphics[scale=0.5]{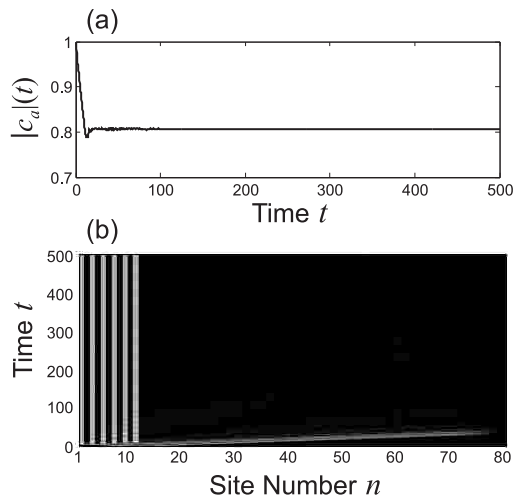} \caption{
Same as Fig.4, but for $\omega_a=0$, corresponding to the
existence of one BIC.}
\end{figure}
Therefore in the tight-binding realization of the colored
Fano-Anderson model BIC arise due to a trapping effect which
localizes the excitation between the boundary of the semi-infinite
chain and the $|n_0-1 \rangle$-th site of the chain. The coupling
of the state $|a \rangle$ with the lattice site $|n_0 \rangle$
allows for the vanishing of the amplitudes $c_n$ at lattice sites
$n \geq n_0$ via quantum destructive interference. It should be
noted that similar trapping mechanisms supporting BIC in
tight-binding models have been recently found in triple or
quadruple dot molecules connected to two leads
\cite{Ladron06,Voo06}, which are modeled as two semi-infinite
tight-binding lattices. In these models, BIC correspond to
vanishing of the wave function at the sites in the molecule in
contact with the leads, i.e. BIC are fully localized in the
molecule sites but not in the leads. Conversely, the present
tight-binding model [Eq.(36)] involves solely one localized state
side-coupled to a semi-infinite lattice, and therefore BIC can not
simply correspond to a decoupling of the localized state with the
lattice. This is clearly demonstrated by the fact that the wave
function of a BIC for the model expressed by Eq.(36) is
non-vanishing {\it even} in a portion of the lattice (from the
boundary site $|1\rangle$ to the site $|n_0-1\rangle$).
Additionally, BIC cease to exist as $n_0 \rightarrow
\infty$, i.e. lattice truncation is essential to sustain BIC.\\
 We checked the existence of BIC
induced by this trapping mechanism by a direct numerical analysis
of the coupled mode equations (33-35) using a fourth-order
variable-step Runge-Kutta method with smoothly absorbing boundary
conditions at the right boundary of the lattice to avoid spurious
reflections due to truncation of Eq.(33). As an example, Figs.4
and 5 show the decay dynamics of $|c_a(t)|$ for parameter values
corresponding to a complete decay (Fig.4), i.e. to the absence of
a BIC, and to a fractional decay (Fig.5) related to the existence
of a BIC. In the figures, the dynamical evolution of the lattice
site amplitudes $|c_n(t)|$ is also depicted on a gray-scale plot,
showing either a diffusion [Fig.4(b)] or a localization [Fig.5(b)]
of the excitation transferred from the localized state $|a\rangle$
to the lattice site $|n_0\rangle$. We checked that the
numerically-computed decay law for $c_a(t)$ exactly reproduces the
curve predicted by the analytical decay law Eqs.(24-26). Note
that, as the resonance condition (26) for the existence of BIC is
satisfied, fractional decay of $c_a$ is attained [Fig.5(a)], which
clearly corresponds to trapping of the excitation at the lattice
sites $ |1\rangle, |2\rangle, ..., |n_0-1\rangle$ with a
destructive interference of site excitation for $n \geq n_0$
[Fig.5(b)]. Conversely, for a frequency $\omega_a$ which does not
satisfy the resonance condition (23) for some integer $m$, the
amplitude $c_a(t)$ fully decays toward zero [Fig.4(a)] and the
excitation transferred to the lattice diffuses along the lattice
without being trapped [Fig.4(b)].

\section{Conclusions}
In this work an exactly-solvable single-level Fano-Anderson model
which admits of bound states inside the continuum has been
proposed, and its relevance to tight-binding lattice models
generally adopted to study electron or photon transport phenomena
in condensed-matter or photonic systems has been discussed. As
previously proposed models supporting BIC in single-level
Fano-Anderson models require point-like gaps in the density of
states \cite{Lambropoulos00,Kilin93,Nabiev93}- a condition which
does not seem to have simple physical realizations
\cite{Lambropoulos00}- in the present work it has been shown that
BIC can exist within a tight-binding continuum without point-like
gaps provided that the interaction of the localized state with the
continuum shows a colored interaction with zero-points. A lattice
realization for such a colored Fano-Anderson model, which may be
of relevance to model photon or electron transport in certain
photonic or condensed-matter systems
\cite{Tanaka06,Stefanou98,Yariv99,Bayindir00,Fan04,Kivshar05,Longhi06,Fiori06},
has been proposed, and a simple physical explanation of the
existence of BIC as a trapping effect sustained by lattice
truncation has been highlighted.


\begin{thebibliography}{99}

\bibitem{vonNeumann29}
J. von Neumann and E. Wigner, Phys. Z. {\bf 30}, 465 (1929).


\bibitem{Stillinger75}
F.H. Stillinger and D.R. Heerick, Phys. Rev. A {\bf 11}, 446
(1975).

\bibitem{Stillinger77}
F.H. Stillinger, Physica B {\bf 85}, 270 (1977).

\bibitem{Herrick77}
D.R. Herrick, Physica B {\bf 85}, 44 (1977).

\bibitem{Friedrich85}
H. Friedrich and D. Wintgen, Phys. Rev. A {\bf 31}, 3964 (1985);
{\bf 32}, 3231 (1985).

\bibitem{Timp88}
G. Timp, H.U. Baranger, P. de Vegvar, J.E. Cunningham, R.E.
Howard, R. Behringer, and P.M. Mankiewich, Phys. Rev. Lett. {\bf
60}, 2081 (1988).

\bibitem{Schult90}
R.L. Schult, H.W. Wyld, and D.G. Ravenhall, Phys. Rev. B {\bf 41},
12760 (1990).

\bibitem{Capasso92}
F. Capasso, C. Sirtori, J. Faist, D.L. Sivco, S.-N. G. Chu, and
A.Y. Cho, Nature (London) {\bf 358}, 565 (1992).

\bibitem{Ji92}
Z.-L. Ji and K.-F. Berggren, Phys Rev. B {\bf 45}, 6652 (1992).

\bibitem{Deo94}
P.S. Deo and A.M. Jayannavar, Phys. Rev. B {\bf 50}, 11629 (1994).

\bibitem{Weber94}
T.A. Weber, Solid State Commun. {\bf 90}, 713 (1994).

\bibitem{Kim99}
C.S. Kim, A.M. Satanin, Y.S. Joe, and R.M. Cosby, Phys. Rev. B
{\bf 60}, 10962 (1999).

\bibitem{Nockel99}
J.U. N\"{o}ckel, Phys. Rev. B {\bf 46}, 15348 (1992).


\bibitem{Olendski02}
O. Olendski and L. Mikhailovska, Phys. Rev. B {\bf 66}, 035331
(2002).

\bibitem{Bulgakov02}
E.N. Bulgakov, P. Exner, K.N. Pichugin, and A.F. Sadreev, Phys.
Rev. B {\bf 66}, 155109 (2002).


\bibitem{Cederbaum03}
L.S. Cederbaum, R.S. Friedman, V.M. Ryaboy, and N. Moiseyev, Phys.
Rev. Lett. 90, 13001 (2003).

\bibitem{Ladron03}
M.L. Ladron de Guevara, F. Claro, and P.A. Orellana, Phys. Rev. B
{\bf 67}, 195335 (2003).

\bibitem{Sprung03}
D.W.L. Sprung, P. Jagiello, J.D. Sigetich, and J. Martorell, Phys.
Rev. B {\bf 67}, 085318 (2003).

\bibitem{Rotter05}
I. Rotter and A.F. Sadreev, Phys. Rev. E {\bf 71}, 046204 (2005).

\bibitem{Ladron06}
M.L. Ladron de Guevara and P.A. Orellana, Phys. Rev. B {\bf 73},
205303 (2006).

\bibitem{Ordonez06}
G. Ordonez, K. Na, and S. Kim, Phys. Rev. A {\bf 73},
022113(2006).

\bibitem{Voo06}
K.-K. Voo and C.S. Chu, Phys. Rev. B {\bf 74}, 155306 (2006).

\bibitem{Fano}
U. Fano, Phys. Rev. {\bf 124}, 1866 (1961).

\bibitem{Anderson}
P.W. Anderson, Phys. Rev. {\bf 164}, 41 (1961).

\bibitem{Mahan}
G.D. Mahan, {\it Many-Particle Physics} (New York, Plenum Press,
1990), pp.272-285.

\bibitem{Friedrichs}
K.O. Friedrichs, Commun. Pure Appl. Math. {\bf 1}, 361 (1948).

\bibitem{Lee}
T.D. Lee, Phys. Rev. {\bf 95}, 1329 (1954).

\bibitem{Prigogine92}
I. Prigogine, Phys. Rep. {\bf 219}, 93 (1992).

\bibitem{Tannoudji}
C. Cohen-Tannoudji, J. Dupont-Roc, and G. Grynberg, {\it
Atom-Photon Interactions} (Wiley, New York, 1992).

\bibitem{Knight90}
P.L. Knight, M.A. Lauder, and B.J. Dalton, Phys. Rep. {\bf 190}, 1
(1990).

\bibitem{Lambropoulos00}
P. Lambropoulos, G.M. Nikolopoulos, T.R. Nielsen, and S. Bay, Rep.
Prog. Phys. {\bf 63}, 455 (2000).

\bibitem{Gadzuk00}
J.W. Gadzuk and M. Plihal, Farady Discuss. {\bf 117}, 1 (2000).


\bibitem{Miyamoto05}
M. Miyamoto, Phys. Rev. A {\bf 72}, 063405 (2005).


\bibitem{Lewenstein88}
M. Lewenstein, J. Zakrzewski, T.W. Mossberg, and J. Mostowski, J.
Phys. B: At. Mol. Opt. Phys. {\bf 21}, L9 (1988).

\bibitem{Piraux90}
B. Piraux, R. Bhatt, and P.L. Knight, Phys. Rev. A {\bf 41}, 6296
(1990).

\bibitem{John90}
S. John and J. Wang, Phys. Rev. Lett. {\bf 64}, 2418 (1990).

\bibitem{John94}
S. John and T. Quang, Phys. Rev. A {\bf 50}, 1764 (1994).

\bibitem{Kofman94}
A.G. Kofman, G. Kurizki, and B. Sherman, J. Mod. Opt. {\bf 41},
353 (1994).

\bibitem{Gaveau95}
B. Gaveau and L.S. Schulman, J. Phys. A: Math. Gen. {\bf 28}, 7359
(1995).

\bibitem{Kilin93}
S. Ya. Kilin and D.S. Mogilevtsev, Opt. Spectrosc. {\bf 74}, 579
(1993).

\bibitem{Nabiev93}
R.F. Nabiev, P.Yeh, and J.J. Sanchez-Mondragon, Phys. Rev. A {\bf
47}, 3380 (1993).

\bibitem{Tanaka06}
S. Tanaka, S. Garmon, and T. Petrosky, Phys. Rev. B {\bf 73},
115340 (2006).

\bibitem{Stefanou98}
N. Stefanou and A. Modinos, Phys. Rev. B {\bf 57}, 12127 (1998).

\bibitem{Yariv99}
A. Yariv, Y. Xu, R.K. Lee, and A. Scherer, Opt. Lett. {\bf 24},
711 (1999).

\bibitem{Bayindir00}
M. Bayindir, B. Temelkuran, and E. Ozbay, Phys. Rev. B {\bf 61},
R11855 (2000).

\bibitem{Fan04}
M.F. Yanik and S. Fan, Phys. Rev. Lett. {\bf 92}, 083901 (2004).

\bibitem{Kivshar05}
A.E. Miroshnichenko and Y.S. Kivshar, Phys. Rev. E {\bf 72},
056611 (2005).

\bibitem{Longhi06}
S. Longhi, Phys. Rev. Lett. {\bf 97}, 110402 (2006).

\bibitem{Fiori06}
E. Rufeil Fiori and H.M. Pastawski, Chem. Phys. Lett. {\bf 420},
35 (2006).

\end{thebibliography}
\end{document}